\documentclass[12pt]{article}

\usepackage{amssymb}
\usepackage{amsmath}

\begin{document}
\newcommand{\md}{misfit dislocation}
\newcommand{\td}{threading dislocation}

\title {\bf Strain relaxation models}
\author{A.I. Zhmakin\\
ai@zhmakin.ru\\
Soft-Impact. Ltd, {Saint-Petersburg, 194156, Engelsa av. 27}}
\maketitle
\vspace{1truecm}
\setcounter{secnumdepth}{4}
\setcounter{tocdepth}{4}
\tableofcontents

\newpage
\section{Misfit strain accomodation}
The lattice mismatch for the semiconductor layer  with lattice
constant $a$ is measured by the misfit parameter $f_m$
\begin{displaymath}
f_m = \frac{a-a_s}{a_s}
\end{displaymath}
where $a_s$ is the lattice constant of the substrate.
The misfit parameter for the $\mathrm{Ge_xSi_{1-x}/Si}$ heterostructure
could be written, using Vegard rule
$a(x) = x a_{\mathrm{Ge}} + (1-x)  a_{\mathrm{Si}}$, as
$f_m = 0.418 x$.

If the  thickness $h$ of the layer is small,
the misfit between the two semiconductors is accommodated by a strain
of the layer that is known as the `misfit
strain'. The in-plain ($x-y$) components of the strain tensor are
\begin{displaymath}
\varepsilon_{xx} = \varepsilon_{yy} = \varepsilon_\parallel = f_m
\end{displaymath}
while the normal one
\begin{displaymath}
\varepsilon_{zz} = \varepsilon_\perp = -2 \frac{C_{13}}{C_{33}} \varepsilon_\parallel
= - \frac{2 \nu}{1 - \nu} \varepsilon_\parallel
\end{displaymath}
where $C_{ij}$ are the components of the elastic stiffness tensor
in Voigt notation \cite{nye} and $\nu$ is the Poisson ratio
that can differ significantly from its bulk value for very thin
films \cite{morozov}.
Subscript $\parallel$ will be omitted below.

The general definition of the elastic energy is
\begin{displaymath}
E^{el} = \displaystyle \frac{1}{2} \mathcal{E} \cdot  \mathcal{C} \cdot \mathcal{E} .
\end{displaymath}
The elastic energy $E$ per unit area stored in the layer
due to the homogeneous strain is \cite{jain91}
\begin{displaymath}
E^{el} = B h f_m^2
\end{displaymath}
where the constant $B$ is defined as
\begin{displaymath}
B = 2 \mu \frac{1+ \nu}{1 - \nu}
\end{displaymath}
and $\mu$ is the elastic shear modulus.
$B$ is not the bulk modulus: it allows for the vertical relaxation of the layer
which accompanies the constraint in the plane, and incorporates the factor of
1/2 in relating elastic energy to the square of strain.

The strain energy $E^{el}$ is proportional to $h$.
As $h$ increases
and exceeds a certain critical thickness $h_c$,
pseudomorphic growth of the uniform layer with flat free surface is no more possible
and several phenomena are observed:
\begin{itemize}
\item{introduction of misfit dislocations}
\item{modulation of the free surface profile}
\item{composition modulations \cite{decomp,spin1,spin2}}
\item{microtwin formation \cite{twin1,twin2}}
\item{interdiffusion between the layer and the substrate \cite{interd,interdif}}
\end{itemize}
The last mechanism  usually occurs at temperatures higher
than typical growth ones.
The composition nonuniformities and microtwin formation
are of primary interest for III-V ternary compounds when one
of the constituents is In. Ref. \cite{tao} lists reasons to
ignore concentration fluctuations in an analysis of SiGe
morhology, the first one being a well-known fact that
Si and Ge are miscible over the entire composition range.
Thus models of the last three mechanisms
are not discussed below.

Strain relaxation through {\md}s and through
the surface modification are certainly the
major routes for the strain accomodation in SiGe alloy.
As a rule of thumb,
the first one dominates for low while
the second - for high lattice mismatch.
However, these mechanisms, depending
on the materials system, growth temperature and
the value of the misfit strain could be cooperative
as well as competitive \cite{hull,desj,hull2}.
On the one hand, there is a direct correlation between
the surface cross-hatched morphology and the arrangement
of the interfacial misfit dislocations \cite{yastr};
surface undulations, in turn, could serve as
nucleation sites. On the other hand, strain relief
by one of this mechanism reduces the driving force
for the other.

The strain, surface and interface energies of the
SiGe/Si heterostructure with and without
misfit dislocations have been recently
computed for all three growth modes
(Frank-van der Merve, Stranski-Krastanov and
Volmer-Weber) as a function of the layer composition
and thickness \cite{nak}.

Introduction of {\md}s could be explained
by an analysis of the energy of the system. For $h > h_c$
introduction of dislocations becomes
energetically favourable providing a partial
strain accommodation. They are introduced
at the interface in the case of the constant
layer composition and throughout the strained
graded layer (uniformly when the linear grading
is used and non-uniformly for the square-root or
parabolic profile \cite{bosa}).

Misfit dislocations could be produced by
the motion of the threading dislocations
from the substrate or generated in the strained layer
via nucleation and/or multiplication. Three stages
(regimes) of the strain relaxation through introduction
of misfit dislocations can be distinguished \cite{gonz,rodr}.
The first one is characterised by a relatively slow
strain relief provided mainly by the glide of the
pre-existing threading dislocations. The relaxation
rate in the second one is higher and depends
on the multiplication processes and activization
of new nucleation mechanisms. In the last stage
a saturation of relaxation is observed caused by
the strain hardening ('work-hardening') \cite{gonz2}.

On the other hand, the elastic energy of a body
with a flat surface {\em always diminishes} if
the surface becomes wavy and thus counteracts the
effect of increasing surface energy. Thus the
strained flat surface could be unstable and
development of surface undulations could relax strain.
The wavelenth $\lambda$ of the surface ripples
is decrease with the layer strain \cite{tham}: $\lambda \propto \varepsilon^{-2}$.
The strain is reduced locally at the peaks of the structure
and  is increased in the throughs.

An extreme stage of surface roughening is the
formation of epitaxial islands that are a promising
object for electronic devices \cite{freund}.
This problem had gain a lot of attention recently
(see, for example,
reviews \cite{island,island1,island2}).
The reverse phenomenon - strain relaxation by pit formation
in the compositionally graded SiGe thick films - also has been
observed \cite{gaspare,gasp2}. Island coalescence could lead
to the formation of the crystallographic tilt due to the
asymmetric generation of $60^o$ dislocation and asymmetric
strain relief \cite{riesz}. It is believed that in contrast to
InGaAs strained layers that are characterized by an instability
against the simultaneous perturbation of the surface profile
and the composition, the onset of the surface roughening of
strained  the SiGe layers is primarily determined by nucleation
of islands \cite{leon}.
Surface roughening is certainly an evil,  if
the aim is to grow a planar layer.

%\newpage
\section{Dislocation system in equilibrium}

Two theories have been developed to calculate the equilibrium critical thickness
$h_c$ of the uniform epitaxial layer.
The first theory originated in the work of Frank and Van der
Merwe \cite{f_merve} and has been developed further by Van der Merwe and collaborators
\cite{merve}.
It is based on the principle of  the energy minimization.
The second one by Matthews and Blakeslee \cite{matt_b,matt}
is known as the force balance theory.
Being correctly formulated, these two theories are equivalent and
give  the identical critical thickness, as by definition of thermodynamic
equilibrium they must. It has been shown that the  expression for
the critical thickness could be also used for graded layers
if the misfit parameter is based on the average Ge concentration
\cite{jain91a}.
Subsequent development of critical thickness models has
been aimed at the accurate modelling of the dislocation core
energy \cite{beltz}, accounting for the surface effects
\cite{cam} and anisotropy \cite{anis}.

If the thickness of the layer is continuously increased, the energy
minimization predicts that the number of {\md}s and the strain relaxation will also increase. The strain is never fully relaxed for any finite value of
the thickness but approaches $f_m$
(which corresponds to the complete relaxation of strain)
as h tends to $\infty$. To calculate the number of dislocations
as a function of $h$,
a minimum of the total energy of the layer
should be determined.

The possible orientations of the {\md}s are limited by
the crystallography of the system.
For the f.c.c. structures of SiGe alloys with the interface
normal coinsiding with a cube edge dislocations form in
two parallel arrays with members of one array being
perpendicular to the members of the other.
Let the spacing between two neighbouring dislocations in the arrays be $p$
and
\begin{displaymath}
b_1 = - b \sin (\alpha) \sin (\beta),
\end{displaymath}
where $b$ is  Burgers vector,
$\alpha$ is the angle between the glide plane and the normal to the interface
and $\beta$ is the angle between the dislocation line and the Burgers
vector \cite{jain92,jain93}.
For $60^o$ dislocations
\begin{displaymath}
\alpha = \arctan \frac{1}{\sqrt {2}}, \qquad \beta = \frac{\pi}{3}
\end{displaymath}
while for $90^o$ dislocations
\begin{displaymath}
\alpha =  \frac{\pi}{2}, \qquad \beta = \frac{\pi}{2}
\end{displaymath}
The in-plain component of the homogeneous strain in the layer
in the presence of dislocations  becomes
\begin{displaymath}
\varepsilon = f_m + \frac{b_1}{p}
\end{displaymath}
and the energy
\begin{displaymath}
E = B h \left( f_m + \frac{b_1}{p} \right)^2.
\end{displaymath}
$f_m$ and $b_1/p$  always have opposite
signs and the homogeneous energy is reduced by the {\md}s
(misfit-energy-increasing dislocations studied in ref. \cite{increase}
are nonequilibrium ones).
The energy of dislocations $E^{\mathrm{d}}$
is determined using linear elasticity (for example, \cite{lan,teo}).
It also contributes to the total energy
that for uniform distribution of dislocations is written as
\begin{displaymath}
E^{\mathrm{tot}} = B h \left( f_m + \frac{b_1}{p} \right)^2  + \frac{2}{p} E^\mathrm{d}.
\end{displaymath}
In the early works the following expression for the dislocation energy has been used
\begin{equation}
\label{ed}
E^\mathrm{d}_\infty = A b^2 \left( (1 - \nu \cos^2 \beta)
\left( \ln\frac{\varrho h}{b}  + 1\right) \right)
\end{equation}
where $A = \mu /(4 \pi (1 - \nu))$  and parameter $\varrho$ accounts
for the non-elastic energy of the dislocation core.
Note that the energy only weakly depends on the concrete value of
$\varrho$ for $h \gg b$.

A number of
both explicit and implicit assumptions
have been made in derivation of this relation and folowing from it
equation for the critical
thickness
\begin{equation}
\label{hc}
h_c=\displaystyle\frac{b(1-\nu\cos^2\beta)\left(
\ln\displaystyle\frac{\varrho h}{q}  + 1\right)
+\displaystyle\frac{8\pi(1-\nu^2)s\gamma}{Bb(1-\nu)}}
{8\pi(1+\nu)\sin\beta\sin\alpha
\left( f_m-\displaystyle\frac{2}{B}\displaystyle\frac{\tilde{\sigma^{\mathrm{fault}}}}
{b\cos(2\alpha)\sin\beta} \right)}
\end{equation}
Often the step energy and the stacking fault energy in eq. (\ref{hc}) are
omitted.

One of the simplifying assumption in the derivation of eqs. (\ref{ed},\ref{hc})
is the neglection of  the interaction between dislocations.
Accounting for this effect leads to \cite{jain}
\begin{displaymath}
E^\mathrm{d} = A\left[ a_0+a_1\ln \left(p
\frac{1-\exp (-g)}{2 \pi q} \right) +
a_2 \frac{g \exp(-g)}{1 - \exp(-g)} -
a_3 \frac{g^2 \exp(-g)}{(1-\exp(-g)^2} -a_2 \right]
\end{displaymath}
where
\begin{displaymath}
a_0=(b_1^2+b_2^2) \left(\sin^2{\alpha}-
\frac{1-2\nu)}{4\pi(1-\nu)} \right), a_1= b_1^2+b_2^2) +(1-\nu)b_3^2,
\end{displaymath}
\begin{displaymath}
a_2=b_1^2-b_2^2, a_3=\frac{1}{2}(b_1^2+b_2^2), g=4\pi\frac{h}{p},
b_2=b\cos\alpha\sin\beta, b_3=-b\cos\beta.
\end{displaymath}
Recently it has been claimed, however, that this expression overestimate
the effect of dislocation interactions \cite{wies2}.

The relations displayed so far ignore the presence of the free surface.
This drawback has been eliminated in ref. \cite{fisher2} using the image
method. However, in this work, in turn, it has been assumed implicitly that
the substrate has the infinite thickness. This restriction seems to be too
severe nowadays due to the importance that have gained so called
{\em compliant\/}  substrates. The problems related to the use
of such substrates ("strain partitioning", critical thickness
reduction) have been analyzed recently in ref. \cite{kast}.

The most rigorous  analysis of the critical thickness
is probably presented in ref. \cite{lee}
where the finite thickness of both  the substrate and  the epitaxial layer as well as
the difference in mechanical properties are taken into account.

In the capped layers relaxation occurs by the introduction of
dislocation dipoles (the expression for the dipole energy
could be found in ref. \cite{jain}). When the cap layer thickness
is less than a certain thickness, a mixture of the single and
paired misfit dislocations has been observed \cite{jin}.

The regular periodic distribution having the lowest energy
is rarely occurs in real systems: the dislocations frequently nucleate
at regenerative heterogeneous sources (defects,
impurities, ledges etc.), and hence form bunches.
Presumably, these bunches are distributed
in a random manner in the layer. For example, statistically
significant measurements of ref. \cite{stat} reveal that
distribution of spacings, being a broad unimodal one at
the begining of  the strain relaxation, could tend to a bimodal
distribution as  the misfit relief proceeds ( the mean spacing
decreases) while in \cite{wies} only significant narrowing
of the unimodal distribution has been registered.

The energy of the non-periodic dislocation arrays has been considered in
\cite{jain93a}.
The total energy of a layer containing non-periodic arrays can be calculated by adding the
homogeneous misfit strain energy and the interaction energy between the homogeneous misfit
strain and the average strain caused by the dislocation arrays.
In equilibrium the number of {\md}s in the layer is smaller if the
distribution is non-periodic.

The primary use of the expression for the total energy of the
layer containing dislocations, as was indicated, is to determine
the critical thickness at which dislocations should appear.
However, one can also get the concentration of dislocation $1/p_e$
that cause strain relaxation $|b_1/p_e|$ and thickness $h_e$ for the equilibrium 'supercritical' layers ($h > h_c$) \cite{jain92}.
The values $|b_1/p_e|$ that provide energy minimum
increase with h first rapidly and then slowly. For any given thickness,
the concentration
of dislocations  $1/p_e$ is smaller if interactions of dislocations
are not properly taken into account. The observed concentrations are always much smaller than
the predicted values for a periodic distribution.
The discrepancy is partly due to the non-periodic distribution and partly
due to the difficulty in nucleating the dislocations.

The effect of the finite size of the substrate or mesa
on the dislocation density reduction \cite{xiong} has been considered
analytically in ref. \cite{fisher} where the distribution of
the misfit stress versus the distance the edge has been obtained
and using finite element method in ref. \cite{anan}.
Extension of the equilibrium theory of the critical thickness
for the epitaxial layers suggested in ref. \cite{huang}
is based on the proper account of the multiple reflection
of the image  dislocations.

%\newpage
\section{Evolution of dislocation system}

In thick ($h > h_c$) semiconductor layers grown at low temperatures
the concentration of {\md}s is much smaller than that
predicted by the thermodynamic equilibrium condition.
Therefore the layers are metastable. When the metastable layers are heated at higher temperatures or during the continuous growth of the layers, dislocations are introduced
and the strain relaxes. Generation of dislocations involves
nucleation and/or multiplication and the glide motion of  the dislocations.
The creation of
 the {\md}s by multiplication also involves  the glide of  the dislocations and the nature of
the dislocations depends on the growth mechanism of the layer.

At high temperatures, the growth mode is by three dimensional island growth
because the atoms can more easily migrate to the islands.
3D growth mode is also occurs in  the high lattice mismatch growth.
In SiGe system with Ge content
over 0.8 three growth stages have been observed \cite{koch}: 1) the pseudomorphic growth
of thick (3-5 ML) wetting layer; 2) nucleation and growth of 3D islands;
3) coalescence of islands and continuos film growth.
Misfit dislocations are readily nucleated at the boundaries between the islands
\cite{payne}.

To develop a model of strain relaxation through the system of dislocations,
it is necessary to describe  the dislocation motion and  the evolution
of the disclocation density due to their primary generation
({\md} forming due the motion of existing {\td}, homogeneos
and/or heterogeneous nucleation),
multiplication, and interactions between dislocations (blocking, mutual
fusion and annihilation)
as well as with  the native and artificially induced (such as
cavities  produced by He or H implantation and annealing
\cite{foll,herzog,trink}) defects.
Development of the dislocation system in the substrate, generally speaking,
should also be taken into account, since the dislocation half-loops in the
substrate could produce a great number of intersections in the glide
planes \cite{stein}. Evidently, the detailed description of the strain relaxation
in the heterostructures is extremely complex and, probably, excessive for
the practical aim of optimization of the growth process.

\subsection{Propagation of dislocations}

The propagation of dislocations at low temperatures is dominated
by a glide; a climb component
that implies mass transport by diffusion in the bulk
is significant at high temperatures only \cite{tsao}.
Velocities of the dislocations of different types can vary greatly.
The thermally activated dislocation velocity is given by the equation
\begin{equation}
\label{vel}
v_d = v_0 (\sigma_{\mathrm{exc}})^m \exp(- E_v/kT)
\end{equation}
where $v_0$ is a constant, $\sigma_{\mathrm{exc}}$ is the excess stress
and $E_v$ is the energy of
activation for the glide motion of the dislocation,
m usually taken as 1 or 2.
The excess stress can be written as
\begin{equation}
\label{exc}
\sigma_{\mathrm{exc}} = 2 S \mu \frac{1 + \nu}{1 - \nu} \varepsilon -
\frac {\mu b \cos (\alpha) (1 - \nu \cos^2 \beta)}
{4 \pi h (1 - \nu)} \ln \frac {\varrho h}{b}
\end{equation}
where $S$ is the Schmid factor. The first term in
Eq. (\ref{exc}) is the stress acting on the dislocation line due to misfit strain and the second term is the self-stress of the dislocation line \cite{f_hull}.

The stress $\sigma_{\mathrm{exc}}$  in the capped layers (strained buried
layers) is as follows \cite{tsao,nix}
\begin{equation}
\sigma_{\mathrm{exc}} = 2 S \mu \frac{1 + \nu}{1 - \nu} \varepsilon -
\frac {\mu b \cos (\alpha) (1 - \nu \cos^2 \beta)}
{4 \pi h (1 - \nu)} \ln \frac {\varrho h}{b}-
\frac {\mu b \cos (\alpha) (1 - \nu \cos^2 \beta)}
{4 \pi h_{\mathrm{eff}} (1 - \nu)} \ln \frac {\varrho h}{b} \nonumber
\end{equation}
where
\begin{displaymath}
h_{\mathrm{eff}}=\displaystyle\frac{h h_{\mathrm{cap}}}{h+h_{\mathrm{cap}}}
\end{displaymath}
Evidently, both $\sigma_{\mathrm{exc}}$  and the velocity of
the dislocation $v_d$ are  smaller in the capped layers.
If the strained layer has an `infinitely thick' capping layer
$h_{\mathrm{cap}}=\infty$, the self-energy of the dislocation line (dipole) increases by a factor 2 \cite{jane92b}
and in the denominator of the second term in the eq.~(\ref{exc}) "2" appears
instead of "4".

The velocity of dislocations in the different regions of a sample
have been observed to be different by a factor up to 3 \cite{hull91}
due to the local variations of  the stress.

The dislocation motion is usually described by  the double (single) kink model.
If the dislocation line is sufficiently long, several kinks
may be formed at the same time.
Single kinks are formed in the thin uncapped layers. As the layer thickness
increases, the rate of nucleation of double kinks also increases.
The transition thickness over which double kinks dominate has been
estimated as 1 $\mu \mathrm m$ and 20 $\mathrm{nm}$ for the
strains $\varepsilon = 0.2 \%$ and $\varepsilon = 1 \%$, respectively
\cite{jain}.
The activation energy is $E_v = E_m + F_k$ for single kink and $E_v = E_m + 2 F_k$
for double kink models,
where $E_m$ is the activation energy for the kink jump along
the dislocation line direction and $F_k$ is the energy required
to nucleate an isolated single kink.

The model \cite{hull91} predicts the linear dependence of the velocity
on the excess stress $\sigma_{\mathrm{exc}}$ and on the length
of the dislocation length when the latter does not exceed
a critical value.

The stability of the dislocation glide has been studied in ref. \cite{lunin}.
The kink motion in the field of random forces has been considered.
It has been found that in the case of  the low stress (compared to the Pierls stress)
the attachment of the point defects to the dislocation core may cause both dislocation
immobilization and instability of the dislocation glide. On the other hand,
experimental data \cite{point} show that the presence of  the point defects could
cause either increase or decrease of the dislocation velocity being
dependent the defect nature, energy, concentration and the layer strain.

The authors of the cited paper have also studied the effect of  the free surface
on the dislocation propagation. While no systematic difference between measurement
during growth and after growth has been registered, the dislocation velocity
has been found to increase several times after forming a native oxide on the
surface in the post-growth processing. The most possible explanation suggested
in the paper is that the local stress at the oxide-layer interface can enhance kink
nucleation rates at the surface.

The substrate thickness is finite, thus, there is strain of the opposite sign and much
smaller in magnitude than in the epitaxial layer and threading dislocation
in the substrate move in the opposite direction. As a result a "hairpin" configuration
is formed. It consists of two long arms parallel to the
surface (one at the interface, the other deep in the substrate),
connected by a small threading segment \cite{pichaud}.

\subsection{Nucleation of dislocations}
If the substrate is characterized by a sufficiently high density
of the pre-existing {\td}s, the necessary for the
strain relaxation {\md}s
are produced  by the propagation of  the threading segments.
When the high quality substrates with  the low dislocation
density are used, the strain relaxation could be limited by  the {\md} generation.

Possible sources of {\md}s are:
\begin{itemize}
\item{homogeneous nucleation of half-loops (whole or partial) at the free
surface of the epitaxial layer}
\item{homogeneous nucleation of half-loops  at the
substrate/epilayer interface}
\item{heterogeneous nucleation of complete loops at the nucleation sites in the bulk
of the epilayer}
\item{heterogeneous nucleation of half-loops at the nucleation sites
(point defects at the interface, edges of the islands}
\item{multiplication of dislocations}
\end{itemize}
The homogeneous nucleation of  the dislocation half-loops at the surface of  the semiconductor strained layers can be analysed through  the behaviour of the total energy of
the loop \cite{jain}
\begin{displaymath}
E^{\mathrm{tot}} = E^{\mathrm{loop}} - E^{\mathrm{strain}}
 \pm E^{\mathrm{step}} + E^{\mathrm{fault}}
\end{displaymath}
where $E^{\mathrm{tot}}$ is the self-energy of the semicircular loop of radius $R$,
$E^{\mathrm{strain}}$ is the reduction of the homogeneous strain energy
due to the interaction between the loop and the
misfit strain and $E^{\mathrm{step}} = 2 R s \gamma$ is the energy of the surface step
which is necessarily created ($s = 1$) or destroyed ($s = - 1$)  if the Burgers vector of the dislocation has a vertical (normal to the surface) component.
The last term $E^{\mathrm{fault}} = \tilde {\sigma}^{\mathrm{fault}} (h / \cos \alpha)$
is included in the case of a partial dislocation
only \cite{island}. It represents the energy of the stacking fault or
the antiphase boundary created by the dislocation.

The total energy of the loop increases from 0 for $R = 0$ to a maximum value
$E^{\mathrm{act}}$, the activation energy
for nucleation (which decreases with increase of the misfit $f_m$).
The maximum occurs when $d E^{\mathrm{tot}}/d R = 0$ at $R = R_c$, the critical radius of the
loop. As the radius increases beyond $R_c$,  the loop energy decreases and
 the loop grows
at a rate determined by the velocity of the dislocations until it reaches the interface.
After this, its
threading segments move apart extending the misfit dislocation at the interface.
The critical loop radius and corresponding activation energy are determined by \cite{island}
\begin{displaymath}
R_c = \displaystyle \frac{B (1 - \nu) (1 - \displaystyle
\frac{\nu}{2}) b^2 \left(1 + \ln \left(
\displaystyle \frac{\varrho R_c}{b} \right) \right) + 16 (1 - \nu^2) s \gamma}
{8 \pi (1 - \nu^2) (\sigma b \sin\beta \sin\alpha \cos\alpha -E^{\mathrm{fault}})}
\end{displaymath}
\begin{displaymath}
E^{\mathrm{act}} = R_c s \gamma +
\frac{B R_c (1 - \nu) (1 - \displaystyle
\frac{\nu}{2}) b^2 \left(\ln \left(
\displaystyle \frac{\varrho R_c}{b} \right) -1 \right)}
{16 (1 - \nu^2)}
\end{displaymath}
If the step and stacking fault energies as well as  the logarithmic factors are neglected,
it can be seen that $E^{\mathrm{act}} \propto b^3$, i.e. it is easier to nucleate
partial dislocations that have smaller Burgers vectors.

The rate of nucleation is generally assumed to be
proportional to $\exp(E^{\mathrm{act}}/kT)$.
The estimates show that values of $E^{\mathrm{act}}$
are extremely high and thus
homogeneous nucleation is very unlikely to occur for the
reasonable values of the misfit parameter \cite{jain}.
In most cases the observed values of the activation
energies are much lower and have been attributed
to  the heterogeneous nucleation. It has been known for many years that point defects and their
clusters lower the activation energy dramatically \cite{matt} and
act as efficient sources for the nucleation of dislocations.
Deliberate control of the number of these sources
(by growing an intermediate layer with  the high density of defects or
introducing defects into the substrate)
is widely used in practice \cite{bolh} and
presents an alternative to the classical layer grading
\cite{abraham}.

The authors of ref. \cite{h_bean} argued that $E^{\mathrm{act}}$ in GeSi alloys
should be lowered for the following reasons:
\begin{itemize}
\item{preferential accumulation of Ge near the core in the compressed layers can
substantially reduce the nonelastic core energy}
\item{random fluctuation of Ge concentration results in the activation energy
reduction in the regions where local Ge concentration is high}
\end{itemize}
In \cite{eaglesham} a regenerative source called 'diamond defect' with the
low activation energy is described which is probably a result of
the interstitial precipitation.
Nucleation at  the atomic ledges trapped at the interface between the
substrate and the epitaxial layer has been suggested in ref. \cite{perovic}.
Unfortunately, little theoretical work on  the heterogeneous nucleation in  the semiconductor strained layers has been done, presumably because many unknowns are involved and the process is very complex.

The modulation of  the free surface (surface roughening) can provide regions
(ripple throughs) of  the large stress where the activation barrier for dislocation nucleation
is extremely low \cite{cullis}. It has been shown that its value is proportional
to $\varepsilon^{-4}$ while for the dislocation nucleation by
other mechanisms usually varies as $\varepsilon^{-1}$ \cite{mooney}.
Thus nucleation of dislocations via surface roughening dominates
for the large values of $f_m$.

Note that the half-loop can nucleate at the surface only if $h \geq h_d$, where
$d = R_c \cos \varphi$,
$\varphi$ is the angle between the surface and the normal to the slip plane.

\subsection{Multiplication of dislocations}

The most popular mechanism invoked for the multiplication of dislocations
is the well-known Frank-Read source or its modification. A characteristic feature of
this mechanism is that often  the large dislocation loops extended into
the Si substrate as well as  the dislocation pile-ups several microns
deep are observed \cite{mooney}.
Another multiplication process is the so-called Hagen-Strunk mechanism \cite{hagen}
that operates when two dislocations meet each other at  the right angle.
Multiplication of dislocations by this mechanism has been observed in
both GeSi \cite{eaglesham} and  InGaAs \cite{fitz89} strained layers.
It operates efficiently, however, only if neither the layer thickness nor
the misfit are  large. There are also doubts of the correctness of
observations interpretation as Hagen-Strunk mechanism in some cases
\cite{vdovin}.
Both Frank-Read nad Hagen-Strunk mechanisms lead to  the bunching of the
dislocations with identical Burgers vectors. Filling the area between this bunches
could be promoted by the cross slip process \cite{hohn,pichaud}.

The rate of multiplication is commonly written as
\begin{equation}
%\label{mult}
\left( \frac{d N}{d t} \right)_{\mathrm{mult}} = K N_m v_d
\end{equation}
where $K$ is a breeding factor and $N_m$
is the number of mobile threading dislocations.
A breeding factor is usually considered to be
either constant \cite{jain} $K = K_0$ or proportional
to the excess stress \cite{dodson} $K \propto \sigma_{\mathrm{exc}}$.

A more elaborate expression has been suggested for the
Hagen-Strunk multiplication mechanism \cite{jain}
\begin{equation}
\label{hagen}
\left( \frac{d N}{d t} \right)_{\mathrm{mult}} =
- \frac {(f_m + \varepsilon) N_m v_d}{2 b_{\mathrm{eff}}} P_{\mathrm{mult}}
\end{equation}
where $P_{\mathrm{mult}}$ is the probability that an interaction of
the appropriate type leads to a
multiplication event that depends on the lattice mismatch and the thickness of the layer.

\subsection{Interaction of dislocations}
Dislocation interactions  not only influence the rate at which  the dislocations propagate,
but also can halt  the threading dislocation motion entirely.
Additionally, blocked dislocations can alter  the surface
morphology as well as limit the overall relaxation of technologically important
low-dislocation-density, graded buffer structures.
On the other hand, annihilation of the threading segments could lead to
 the significant reduction of the dislocation density in the strained layer.

\subsubsection{Blocking (pinning) of dislocations}

Two blocking mechanisms are known. The first one, {\em long-range blocking\/},
has been described over a decade ago \cite{freund}. Recently another
blocking mechanism named {\em reactive blocking\/} has been detected both
experimentally (by real time transmission electron microscopy observations)
and numerically (using discrete disclocation dynamics computations of
the strained layer relaxation) \cite{stach}.

\paragraph{Long-range blocking}

The {\md} could impede the motion of  the {\td}  if the two dislocations have the right kind of Burgers vectors. There are four pairs of strain relieving Burgers vectors,
only one of which causes  the significant blocking of
the moving dislocation. The probability that the dislocation interaction can impede the motion of the threading dislocation is therefore 1/4.

As a propagating threading dislocation approaches an interfacial misfit dislocation
segment, the strain fields associated with each dislocation begin to overlap, resulting in an
interaction force that has  the general form:
\begin{displaymath}
\sigma_{\mathrm{int}} \propto \frac {\mathbf{b_1} \cdot \mathbf{b_2}}{r}
\end{displaymath}
where ${\mathbf{b_1}}$ and ${\mathbf{b_1}}$
are the Burgers vectors of the two dislocations and $r$ is the distance between
them. Thus, this force can act to either increase or decrease the magnitude of the net stress
that drives the threading dislocation forward, depending on the signs of the Burgers vectors
of the dislocations involved. If this
interaction stress cancels the other stresses over a significant portion of the threading
segment, the motion of the entire dislocation will be halted. If the interaction stress is not sufficiently large, the threading dislocation will propagate past the misfit interfacial segment.

To bypass the blocking {\md},  the {\td} should alter its path by moving in the same
glide plane but closer to the free surface, i.e. in a channel of the width $h^\star$
smaller than the layer thichness. Three forces act on  the {\td} in this configuration:
\begin{enumerate}
\item{the driving force due to the residual homogeneous strain}
\item{the retaining force due to the line tension of the {\td} (including the
interaction with the surface via the image forces}
\item{the interaction force with the {\md}}
\end{enumerate}
The condition of blocking the {\td} motion could be written as
an equation for critical value of $h^\star$ \cite{russel}
\begin{displaymath}
\varepsilon - \varepsilon_r = \frac{3 b}{16 \pi h^\star (1 + \nu)}
\left(\frac{4 - \nu}{3} \ln \left(\frac{8 h^\star}{b} \right) -
 \frac{1}{2} \cos(\alpha) -  \frac{1 - 2 \nu}{4 (1 - \nu)} \right ),
\end{displaymath}
where $\varepsilon_r$ is the reduced strain due to the presence of {\md}.

The described blocking mechanism is enhanced  due to the presence of the surface
ripples as result of the stress field of the orthogonal misfit
dislocation \cite{sf97,fitz99} and could lead to the {\td}s pile-ups
to be discussed later in this section.

\paragraph{Reactive blocking}

A new strong blocking effect observed using real time TEM has been
studied by numerical simulations \cite{stach}. A propagation of the threading
segment towards  the {\md} on an intersecting glide
plane has been analysed. With the available Burgers vectors and
directions of approach, sixteen distinct interactions of this kind are possible, leading to a
variety of outcomes such as repulsion of the misfit dislocation into the substrate,
reconnection of the two dislocations, and junction or jog creation.
It was found that four of the
interactions involve the  parallel Burgers vectors and can result in a reconnection reaction.
The authors claim that this blocking mechanism is much stronger than the
conventional (long range) misfit-blocking interaction. Unfortunately,
no analytical model/approximation is proposed in the paper.

\paragraph{Overall blocking effect}
The number of dislocations blocked per unit time is defined as \cite{jain}
\begin{equation}
\label{block}
\left( \frac{d N}{d t} \right)_{\mathrm{block}} =
\frac{d N_i (t)}{d t} P(t)
\end{equation}
where $N_i(t)$ is the total number of interactions  and P (t) is the
blocking probability. P (t) is 1/4 if the blocking occur ( the local force is greater
than the critical value) and zero otherwise. Eq. (\ref{block}) could be re-written as
\begin{equation}
%\label{block1}
\left( \frac{d N}{d t} \right)_{\mathrm{block}} =
- \frac {(f_m + \varepsilon) N_m v_d}{2 b_{\mathrm{eff}}} P(t) \\ \nonumber
\end{equation}
If the propagating {\td} interacts with a closely bunched cluster of N {\md}s with  the identical Burgers vectors, the interaction force is multiplied by a factor N \cite{freund90}.

The pile-up formation mentioned above could be responsible for
the strain (work) hardening \cite{hull_bacon} and observed slowing of strain relaxation in
the layer at  the late stages. An indirect confirmation of the importance of
the pile-ups formation in SiGe is the effect of the chemical-mechanical
polishing at the intermediate growth stages on the final {\td} density
reduction \cite{fitz_apl}. It is believed that planarization of the
surface free the {\td}s pinned in the pile-ups.

Introduction of hardening is a way to account integrally for the numerious interactions between
the dislocations.
While on atomic level plastic flow is always very inhomogeous, its macroscopic
phenomenological description via  the flow stress
\begin{displaymath}
\tau= \alpha_\tau \mu b \sqrt {\rho}
\end{displaymath}
where  the coefficient $\alpha_\tau$ depends on the strain rate and the temperature
is commonly used and proved to be reasonable for most cases of strain hardening in
a wide range of materials \cite{kocks}.

\subsubsection{Fusion and annihilation of dislocations}
The outcome of the binary reaction of two threading
dislocations depends on the relative arrangement of their
gliding plabes and  the orientation of the Burgers vectors.
Possible cases including fusion and annihilation of dislocations
have been considered in ref. \cite{pichaud}. Assuming that
the first threading dislocation propagates in the plane
(111) and its Burgers vector $\mathbf b_1$ has a positive projection
on the vertical direction, basic binary reactions are  summarized
in the Table 1.
\begin{table}
\caption{Basic binary reactions between threading dislocations}
\begin{tabular}{llll}
\hline
$2^{nd}$ plane & $(b_2)_z$ & Treading & Misfit\\
\hline
111 & positive & fusion & single line\\
111 & negative & annihilation & single line\\
$1\tilde11$ & positive & fusion & two-arm\\
$1\tilde11$ & negative & annihilation & two-arm\\
\hline
\end{tabular}
\end{table}
The critical parameter for these reactions is the interaction radius.
A continuum-based approach using linear elasticity has been employed
to compute this variable for the dislocations in the
heteroepitaxial system in ref. \cite{anni}

%\newpage
\section{Strain relaxation models}
\subsection{Discrete models}

A number of both micro- and mesoscale numerical models
\cite{phillips} have been
applied to the study of the relaxation mechanisms of
the strained epitaxial layers.

\subsubsection{Atomistic models}
First-principles total energy computations
has been used to resolve the disagreement of the
experimentally determined relation between
lattice relaxation in in-plane and out-of-plane
directions with the predictions of classical
elasticity \cite{oht}. It was found that
segregation at the interface significantly
influence strain relaxation in the heterostructure.

The deformation state of the heteroepitaxial strained system has been
studied using atomistic simulations in \cite{mar1,mar3}. A three-step
relaxation procedure has been developed:
\begin{enumerate}
\item{structural relaxation with composition being fixed}
\item{compositional relaxation}
\item{further local structural relaxation}
\end{enumerate}
Conjugate gradients method has been used for the energy minimization
at the first and third stages while Metropolis implementation
of Monte Carlo method has been applied to the compositional relaxation.

Molecular dynamics simulations have been used in ref. \cite{increase}
to capture the growth process at the atomic level and to study the
mechanisms of  the dislocation formation. The embedded atom method has been employed
that in addition to the binary interactions efficiently accounts for the many-body
effects. The kinetic constrained influence on the atomic assembly process
has been studied.

In ref. \cite{ham}  the two-dimensional Frenkel-Kontorova model
has been applied to computation of the dislocation nucleation rate in
the growing heteroepitaxial island. As in the preceding paper,
the embedded atom method has been used to compute the total energy.

One-dimensional Monte Carlo method has been used to simulate the surface
height evolution during and after the strain relaxation in ref. \cite{andrews}.
The aim of this study was to get insight into the cross-hatch morhology
development and to asses different existing models of the process  (such
as enhanced growth over strain relaxed regions due to the lateral transport
by surface diffusion and surface undulations caused by the dislocation generation
and glide). The authors conclude that surface step flow is a necessary
condition for the development of the mesoscale cross-hatch morphology
while the plastic relaxation itself could not produce the undulations
of significant amplitude.

\subsubsection{Mesoscopic models}
Detailed simulations of the interaction of the two threading segments
encountering each other in a thin strained SiGe layer has been performed
in ref. \cite{schw7} using the full three-dimensional Peach-Koehler formalism
\cite{hirth}. The force acting on a dislocation segmeny $d\mathbf l$ in the glide
plane is
\begin{displaymath}
b_i\sigma_{ij}n_j (\mathbf n \times d\mathbf l)
\end{displaymath}
where $\mathbf n$ is the normal to the glide plane. The stress
tensor includes stresses due to the applied strain and stresses
generated by the presence of dislocations. The authors have avoided
the difficulties with the stress correction caused by the presence of
surfaces by considering the symmetrically capped layer.

In a few recent papers \cite{stach,don,nicola,schw} the application of
the discrete dislocation dynamics method to the strain relaxation
has been reported. Large scale 2D simulations are used to study
the misfit strain relaxation in the hetroepitaxial islands in ref. \cite{don},
pecularities of the hardening in the single crystal thin films has been
investigated in ref. \cite{nicola}.
Monitoring of the evolution of a few hundreds dislocations
in the strained layer \cite{stach,schw} leaded to a
discovery of a new blocking mechanism discussed briefly in the preceeding
section.

\subsection{Continuum models}
A classification of continuum (macroscopic) models adopted below is by no means
unique and generally accepted. Still, it  is worth to make an attempt to
sort out different approaches to the simulation of the strain relaxation.
The main problem is, of course, a considerable overlapping of ideas and methods.

\subsubsection{Equilibrium models}
\paragraph{Uniform layer}
The equilibrium density of the {\md} in the strained layer with the
uniform composition is obtained similar to the analysis of
the critical thickness by the energy minimization \cite{tsao}.
A coarse estimate of the {\td} density as 1-2 times that for {\md}
follows from the scheme of  the {\md} generation due to the
threading segment motion ($\rho_{\mathrm{td}} = \rho_{\mathrm{md}}$)
or by half-loop nucleation ($\rho_{\mathrm{td}} = 2 \rho_{\mathrm{md}}$).
These estimates, of course, do not account for the {\td} density
reduction due to the fusion/annihilation.

An estimate for the {\td} density via the average {\md} length
has been suggested in ref. \cite{fs77}, assuming that two threading
dislocation are connected by a misfit segment with lenghth $\langle  l \rangle$.
Similar relation has been suggested later \cite{ferrari}:
\begin{displaymath}
\rho_{td} \approx 4 \rho_{md} \left(\frac{1}{\langle  l \rangle} - \frac{1}{L} \right)
\end{displaymath}
where L is the sample size and $\rho_{md}$ is determined via the lattice mismatch.

A model for the equilibrium {\td} density in the thick layer (compared to
the critical thickness) has been
analysed in \cite{ayers} with neglection of the misfit strain. The author
assumes that the population of the {\td}s is governed by the coalescence of the
close dislocations and introduces the 'minimum stable separation' (i.e.
the fusion/annihilation radius) estimated as
\begin{displaymath}
\displaystyle\frac{1}{r_{\mathrm{min}}} =
\displaystyle\frac{\cos \phi}{4h} \left(\cos^2\beta +
\displaystyle\frac{\sin^2\beta}{4(1-\nu)}
\ln \left( \displaystyle\frac {\sin\alpha\sin\beta}{4f_m}\right)\right)
\end{displaymath}
Then $\rho_{\mathrm{td}} = 2 /(R_{\mathrm{av}} r_{\mathrm{av}})$ where
$R_{\mathrm{av}}$ is the average spacing between the glide planes defined by
the misfit and $r_{\mathrm{av}} = 2 r_{\mathrm{min}}$ is the average
distance between the {\td}s within the plane. The model reasonably
predict both the mismatch dependence and the order of magnitude
of the {\td} density for some materials in the range $f_m = 0.002-0.1$.
Still, the author, being aware of the simplifications made, lists major of
them: 1)  the large layer thickness $h \gg h_c$; 2) approximations in the line
tension calculations; 2) assumed large spacing of  the {\md}s;
4) no kinetic barriers to the glide of  the {\td}s; 5)  the {\td}s density is
considered as a function of the film thickness only, while
experiments show that it varies across the layer.

The author also notes that  the difference between the strain accomodation by
$60^o$ {\md}s and the pure edge ones (a factor 2 in the {\td}
density) explaines the twofold reduction of the dislocation density
in some materials during the post-growth annealing by transformation of
the first type dislocations into the second one.

\paragraph{Graded layer}
Strain relaxation in the linear graded epitaxial layers has been considered
in ref. \cite{kim}. The term "equilibrium dynamics" used by the authors is
somewhat misleading. In fact a quasi-stationary approach is exploited. A set
of algebraic equations that define the current values of the lattice constant,
strain, biaxial modulus and shear modulus as a function of the film thickness
is formulated. Expressions for the local relaxation thickness
$h_c^l$, the plastic strain
and the equilibrium dislocation density are obtained.
The value of $h_c^l$ is attributed to the size of the dislocation-free
region on the top of the growing layer.
Its predicted weak dependence on the film thickness
as well as the strong effect of the grading rate on both
the local relaxation thickness and the equilibrium disclocation spacing
are confirmed by the experimental data.

The influence of the grading law on the residual strain,
the {\td} and {\md} density has been studied theoretically and
experimentally in ref. \cite{romanato}. The authors assume that the grading,
however, does not change the basic phenomena such as nucleation/multiplication
studied in detail for the uniform layers. The standard relation between
the strain and the {\md} density is generalized to
\begin{displaymath}
\varepsilon(h)=-f_m(h) + b_{\parallel} \int_0^{h_f} \rho_{\mathrm{md}} dh
\end{displaymath}
where $\varepsilon(h)$ and $f_m(h)$ are the depth profiles
of the residual strain and of the lattice misfit, respectively,
$h_f$ is the total film thickness. Assuming full relaxation,
the authors get
\begin{displaymath}
\rho_{\mathrm{md}}= \displaystyle\frac{1}{b_{\parallel}}
\displaystyle\frac{d}{dh}f_m(h)
\end{displaymath}
Accounting for the existence of the top dislocation-free layer
of the thickness $h_c$, the residual strain distribution is written as
\begin{equation}
\label{eps}
\varepsilon(h)= \left \{
\begin{array}{rl}
0, \quad if \quad 0\leq h \leq h_f-h_c \\
-(f_m(h)-f_m(h_f)), \quad if \quad h_f-h_c < h \leq h_f
\end{array} \right.
\end{equation}
Several layers with different grading laws (linear, parabolic, square-root,
step + linear) have been grown. The analysis of the observed work hardening
forced the following modification of eq. (\ref{eps}):
\begin{displaymath}
\varepsilon(h)= \left \{
\begin{array}{rl}
\varepsilon^{\mathrm{wh}}(h),
\quad if \quad 0\leq h \leq h^{\mathrm{wh}} \\
-(f_m(h)-f_m(h^{\mathrm{wh}})-\varepsilon^{\mathrm{wh}}(h)),
\quad if \quad h ^{\mathrm{wh}}< h \leq h_f
\end{array} \right.
\end{displaymath}
where the supersript 'wh' refers to workhardening and,
as experiments show, $h^{\mathrm{wh}} > h_f - h_c$.

\subsubsection{Reaction kinetics models}
An evolutionary approch based on {\em reaction\/} and
{\em reaction-diffusion\/} models has been applied to a number of the misfit strain
relaxation problems. A general form of kinetics equations
used to study the {\td} reduction in the strained layers  is as follows \cite{rom1}-\cite{rom4}
\begin{equation}
\label{react}
\frac{d \rho_i}{d h} = - \sum_j K_{ij} \rho_i \rho_j
  + \sum_l \sum_m K_{lm} \rho_l \rho_m
\end{equation}
where $\rho_i$ is the density of the specific $i^{\mathrm{th}}$ dislocation family
and  the kinetic coefficients $K_{ij}$ are the rates of the reactions between
 the dislocations from the  families $i$ and $j$.
These equations described the dislocation densities evolution
\begin{itemize}
\item{with the layer thickness during growth $\rho_i = \rho_i (h)$ or}
\item{in time for the film of the fixed thickness $\rho_i = \rho_i (t)$}
\end{itemize}

Both first- and second order reactions could be considered (the order of
a reaction corresponds to the number of participants). For example,
blocking of the {\td} propagation due to the interaction with the {\md}
is the first-order reaction while both fusion ($\mathbf{b}_3 =
\mathbf{b}_1 + \mathbf{b}_2$) and annihilation ($0 = \mathbf{b}_1 + \mathbf{b}_2$)
are the second-order reactions.

The complete treatment of  the strain relaxation using  the reaction equations
requires the account of the crystallographic details and a subdivision of the
dislocation system into separate populations corresponding to  the specific Burgers vectors
and line directions. For f.c.c. semiconductors, for example,
there are 24 dislocation sets arising from the combination of four possible (111)
type slip planes and six Burgers vectors;
20 families of dislocations in GaN have been considered in ref. \cite{rom4}.

The coupled system of nonlinear ODEs being rather complex,
reduced models are frequently used.
An obvious   bonus of the model reduction is the
possibility to obtain an analytical solution for
some limiting cases \cite{rom1,rom3}.

For example, in the model of ref. \cite{hull89} {\md}s are
created exclusively by lateral bending of the threading segments
and {\td}s by half-loop nucleation at the surface at rate $j$;
{\td}s are blocked by the misfit ones with the probability $\eta$
and multiplication is neglected. Thus the following system is used
\begin{eqnarray}
\frac{\partial \rho_{md}}{\partial t} & = & v \rho_{td} \\ \nonumber
\frac{\partial \rho_{td}}{\partial t} & = & j - \eta v \rho_{td} \rho_{md} \nonumber
\end{eqnarray}

The review of some early reaction type  models has
been given in \cite{ayers}. The first one ("annihilation-coalescence" model)
is just an equation for the total dislocation density that accounts
for the fusion and annihilation of dislocations
\begin{displaymath}
\frac{d \rho}{d h}= -A \rho - B \rho^2
\end{displaymath}
leading to the relation
\begin{displaymath}
\rho(h)=\displaystyle\frac{1}{\left(\displaystyle\frac{1}{\rho_0}+
\displaystyle\frac{B}{A} \right)\exp(Ah)-\displaystyle\frac{B}{A}}
\end{displaymath}
The "half-loop" model based on the assumption that the fusion
of the threading dislocations results in the formation of half-loops
and half-loops smaller than a certain critical size are removed from
the layer by gliding to the interface leads to the following  equation for
the total dislocation density
\begin{displaymath}
\rho=\displaystyle\frac{f_m \sqrt{2} (1-\nu)(1-2\nu)\ln \left(
\displaystyle\frac{2\pi f_m}{1-\nu}\right)}
{b h(1-\nu)^3 (1-\ln(2b \sqrt{\rho}))}
\end{displaymath}

The set containing three
dislocation families (mobile and immobile {\td}; {\md})
has been exploited in ref. \cite{rom3} and extended in ref. \cite{rom5}
to  the four unknowns system by splitting the population of
 the {\md}s into an 'active' and a 'passive' parts.
Analytic solutions have been obtained for a number of
special cases (no blocking of the {\td} propagation by the
{\md} or no annihilation reaction). Eqs. (\ref{react})
is used to study  the dislocation evolution either in the layer
of  the fixed thickness or during the growth, i.e. for the
single independent unknown.

Dislocation densities of the gliding and climbing
{\td}s as well as of the {\md}s has been considered
in ref. \cite{rom6}. The authors stress the importance of the
climb process inclusion into the model since it permits
the description of the effect of point defects on the
dislocation propagation.

Another feature of this model is the attempt
to account for the nonlocal character of the {\td}
interaction with the {\md} by introducing the diffusion
term into the conservation equation for the gliding
dislocation density. The principal character of this
extension is the transition from ODE to PDE:
\begin{eqnarray}
\frac{\partial \rho_g}{\partial t} & = & A \sigma(z) - B (\rho_g)  \rho_g+
D \frac{\partial^2 \rho_g}{\partial z^2}\\ \nonumber
\frac{\partial \rho_c}{\partial t} & = & B (\rho_g) \rho_g - K \rho_c\\ \nonumber
\frac{\partial \rho_m}{\partial t} & = & K \rho_c \nonumber
\end{eqnarray}
where $\rho_i$, $i = g, c, m$ are the density of the gliding, climbing
and misfit dislocations, respectively, A, B, K - corresponding reaction rates
\cite {rom6,rom7}.

The model just described has been applied to the problem of the misfit dislocation
patterning \cite{rom7,rom8}. The complete system has been used for the linear stability
analysis only, while dislocation evolution has been considered for  the
two limiting cases: the uniform time-dependent solution
$\rho_i = \rho_i (t)$ and the steady-state non-uniform one
$\rho_i = \rho_i (z)$.

\subsubsection{Plastic flow models}
Phenomenological plastic flow models
of  the dislocation evolution are based on the well-known Alexander-Haasen (AH) model
\cite{ah} developed for elemental semiconductors loaded in a single slip
orientation. It uses the dislocation density as a state variable
and relates the plastic deformation in the crystal to the movement
and multiplication of dislocations. Usually under AH model (or
Alexander-Haasen-Sumino model) a tuple
of three components is meant \cite{dupret}. These components are:
\begin{enumerate}
\item{Orowan equation \cite{orowan} that relates the plastic shear strain rate
to the motion of mobile dislocations
\begin{displaymath}
\dot{\varepsilon}^{\mathrm{pl}}= N b v_d
\end{displaymath}
}
\item{normalized expression for the dislocation velocity eq. (\ref{vel})
with a substitution $\sigma_{\mathrm{eff}}$ instead of $\sigma_{\mathrm{exc}}$
\begin{equation}
v_d = v_0 \left(\frac{\sigma_{\mathrm{eff}}}{\sigma_{\mathrm{0}}}\right)^m
\exp \left(- \frac{E_v}{kT} \right) \nonumber
\end{equation}
}
\item{equation for the dislocation density evolution
\begin{displaymath}
\dot{N}= \delta N v_d
\end{displaymath}
}
\end{enumerate}

The main contribution to the model by Alexander and Haasen themselves is the
adaptation of the relation for the dislocation velocity to
the case of covalent crystals with high Peierl barrier.
They gave an explanation to the Arrhenius type temperature dependence
of the dislocation velocity observed in experiments, in particular,
in Ge \cite{ge}. To determine the backstress $\hat{\sigma}$ in
\begin{displaymath}
\sigma_{\mathrm{eff}} = \sigma_{\mathrm{exc}} - \hat{\sigma}
\end{displaymath}
the authors consider a statistical arrangement of $N$ parallel
dislocations giving the backstress
\begin{displaymath}
\hat{\sigma}= \frac{\mu b}{2\pi(1-\nu)} N^{1/2}
\end{displaymath}
that is consistent with the square-root dependence suggested by G.I. Taylor.

The multiplication law is due to Johnson \& Gilman \cite{gil}.
Alexander and Haasen postulated that the breeding coefficient is
\begin{displaymath}
\delta = K \sigma_{\mathrm{eff}}
\end{displaymath}
where K is an empirical constant.

Refs. \cite{dodson,nix,houghton} are probably the first examples of the
application of AH-type models to the strain relaxation in the  thin films.
The main modification of the plastic flow model in the first of the cited
papers is an analysis of both gliding and climbing dislocation motion
resulting in the equation for $\gamma = f_m - \varepsilon$
\begin{displaymath}
\frac{d\gamma}{dt} = \frac{\sigma_{\mathrm{eff}}^2}{\mu^2}
\left( \Gamma_g \exp -\frac{E_g}{kt} + \Gamma_c \exp -\frac{E_c}{kt}
\right) (\gamma + \gamma_0)
\end{displaymath}
or
\begin{displaymath}
\frac{d \ln(\gamma+ \gamma_0)}{dt} = \frac{\sigma_{\mathrm{eff}}^2}{\mu^2}
\left( \Gamma_g \exp -\frac{E_g}{kt} + \Gamma_c \exp -\frac{E_c}{kt}
\right)
\end{displaymath}
where $\Gamma_g$ and $\Gamma_c$ are  the glide and climb prefactors, respectively,
$E_g$ and $E_c$ are  the activation energies and $\gamma_0$ represents a "source"
term needed for starting multiplication process.
As noted in \cite{tsao}, in  the more accurate formulation the excess in-plane
stress should be replaced by the excess stress resolved on the slip plane.
To adjust $\Gamma_g$,  $\Gamma_c$, $E_g$,  $E_c$ and $\gamma_0$,
experimental data \cite{sige1,sige2} have been used.

The essence of the 'improved' Dodson-Tsao model proposed in \cite{jain}
is the account for different processes that change the dislocation density
\begin{displaymath}
\frac{dN}{dt} = - Q_{\mathrm{block}} + Q_{\mathrm{nucl}} +  Q_{\mathrm{mult}}
\end{displaymath}

In ref. \cite{fitz_dyn} AH-type model has been applied to the strain relaxation
in the graded SiGe layer. Under a number of simplifying assumptions
the authors get an expression for time derivative of the dislocation density
that depends linearly on both  the growth rate $R_g$ and  the grading rate $R_{gr}$
\begin{displaymath}
\frac{d \rho}{dt} \propto R_g R_{gr}
\end{displaymath}
Clearly, this relation does not endure the limiting transition to the case
of the uniform layer growth.

In the investigation of the strain relaxation in the uniform layer \cite{wang}
$\sigma_{\mathrm{eff}}$ has been used in the equations for both the dislocation
velocity and dislocation density evolution. However, the backstress has
been determined as
\begin{equation}
\label{hard}
\hat{\sigma} = B H (\varepsilon^{\mathrm{pl}})=
B \alpha \left(\frac{\varepsilon^{\mathrm{pl}}}{f_m} \right)^{\beta}
\left( 1 - \tanh \frac{\gamma \varepsilon^{\mathrm{pl}}}{f_m} \right)
\end{equation}
where $\alpha, \beta, \gamma$ are ajustable parameters.

This model has been extended to the case of multiple and graded layers in ref.
\cite{wang_am}. The source term in the equation for the dislocation density
evolution has been separated into nucleation and annihilation parts
\begin{displaymath}
\dot{\rho} = \dot{\rho}_{\mathrm{nucl}} + \dot{\rho}_{\mathrm{annih}}
\end{displaymath}
A threshold stress for the {\td} nucleation $\sigma_0$ has also been introduced
\begin{displaymath}
\dot{\rho}_{\mathrm{nucl}} = \xi_0 \left(
\frac {\sigma_{\mathrm{exc}} - \hat{\sigma} - \sigma_0}{\mu} \right)
\exp -\frac {E_{\rho}}{kT}
\end{displaymath}
The autors assume that dislocation multiplication plays a relatively minor role;
the threading dislocations nucleate at the surface and distributed to all the layers according to some weight - a power function of the effective stress has been adopted in the paper.

An extension of the Dodson-Tsao model \cite{dodson} suggested in ref. \cite{beres}
is mainly a modification of the rate equation for the surface nucleation.
In addition to the nculeation equation itself
\begin{displaymath}
\frac{d \rho}{d t} = \xi_0 \left(
\frac{\sigma_{\mathrm{eff}}}{\mu}\right)^{(1 + \alpha)}  N_s
\end{displaymath}
an equation for the time evolution of the nucleation site
density $N_s$ is introduced
\begin{displaymath}
\frac{d N_s}{d t} = G - \frac{N_s}{t_0}
\end{displaymath}
where $t_0$ is the characteristic time of the source deactivation.
This modification seems to be specific for III-V heterostructure
growth where, in contrast to the SiGe material system,
a significant reduction of the strain relaxation rate at the growth
interruption is observed.
The extended models gives the deviation from the experimental data
twice smaller than the model \cite{dodson}.

The AH-type model has been used  to study the strain relaxation in
the structure with the substrate of finite thickness in
the multiscale approach of ref. \cite{mar3}.
The model equations have been combined \cite{m_apl,m_ss}
with the equation of the mechanical equilibrium
\begin{displaymath}
M_f \varepsilon_f h_f + M_s \varepsilon_s h_s = 0
\end{displaymath}
and the compatibility equation
\begin{displaymath}
\varepsilon_f - \varepsilon_s = f_m - s b N_{md}
\end{displaymath}
where
\begin{displaymath}
s = \left \{
\begin{array}{rl}
1  \qquad \qquad for \; tensile\; strain \qquad f_m > 0 \\
-1  \qquad for \; compressive \; strain  \qquad f_m < 0
\end{array}\right.
\end{displaymath}
to get an ODE for the single unknown - the strain in the layer $\varepsilon_f$ -
as a function of the film thickness \cite{mar3,m_ss}.

%\newpage
\section{Conclusions}
\subsection{Strain relaxation scenario}
The overall scenario of the strain relaxation in the heteroepitaxial structure
with low/medium lattice mismatch includes the following stages:
\begin{enumerate}
\item{elastic strain accomodation}
\item{slow strain relaxation}
\item{fast strain relaxation}
\item{relaxation saturation due to strain hardening}
\end{enumerate}
An additional process to be considered is  the relaxation during the annealing. This stage
is somewhat simpler in the numerical analysis since the film thickness is fixed.

\subsection{Assessment of strain relaxation models}
\subsubsection{Discrete models}
The advantage of the micro- and mesoscale numerical models of the misfit
strain relaxation is the detailed description of the processes. There are,
however, two drawbacks. The first one is evident: huge computer resources
for the real-life problems.
The second one is more subtle, but actually more severe: the need
for corresponding initial and boundary conditions.

Thus, at the present time, such discrete numerical models could be practically
usefull
\begin{itemize}
\item{as a component of a multiscale simulation system either directly or via homogenization-type procedure}
\item{as a measuring stick for the calibration of the continuum models \cite {mar2}}
\end{itemize}

\subsubsection{Continuum models}
It is evident that estimate of the dislocation density
using equlibrium models will always produce an {\em upper\/} bound
of this parameter and a {\em lower} bound for the residulal strain.

Reaction and reaction diffusion models allow a detailed description of the
interaction of the dislocations belonging to the different slip systems.
Their weak point is first of all the absence of a mechanism to account for the
collective behavour of the dislocations and its effect on the strain.
The known applications of the models of this kind deal with either of
\begin{itemize}
\item{dislocation evolution versus the film thickness}
\item{with annealing of the constant thickness film}
\end{itemize}
The reaction-diffusion models described above that include the gradient
terms are formulated as two dimensional problems. The spatial coordinate is
in-plane, however, and solutions published are either uniform (zero spatial
dimension)  non-stationary or 1D stationary.

Plastic flow models relay heavily on the tuning to the experimental data.
Still, it seems that at present they are capable to account best
(phenomenologically) for the complex processes of the strain relaxation
in the heterostrucures, provided enough experimental information is available
for the reliable determination of the adjustable parameters. In the
all known examples plastic flow models used for the relaxation during
growth are written in terms of the layer thickness as an independent variable.
These models have been also applied to the simulation annealing.

\subsection{Evolutionary model}
The comprehensive model of misfit accomodation should describe the strain relaxation and
the dislocation evolution both during the growth itself and
the post-processing (annealing). As the first step we are considering the layer
only without the substrate.

The model for the strain relaxation in the heterostructure being implemented now
is an adaptation of the model developed for the analysis of dislocation
density evolution during
the growth of single bulk crystals \cite{oxford}.
Its major features are as follows.

The strain is divided into elastic and plastic components:
\begin{displaymath}
\varepsilon_{ik} = \varepsilon^{el}_{ik} + \varepsilon^{\mathrm{pl}}_{ik}
\end{displaymath}

Strain is related to the stress via Hooke's equation which in
general case of anysotropic crystal can be written as
\begin{displaymath}
\sigma_{ik} =
   \sigma^{el}_{ik} +
   c_{iklm} ( \varepsilon^{\mathrm{pl}}_{lm} - \beta_{lm} \triangle T )
\end{displaymath}

where $\sigma_{ik}$ is the stress for completely elastic case
which should be used in the equilibrium equations,
$\sigma^{el}_{ik}$ is the real elastic stress,
$\triangle T$ is relative temperature,
$c_{iklm}$ are elastic constants and
$\beta_{ik}$ are thermal expansion coefficients,
which are usually assumed to be isotropic:
$ \beta_{ik} = \delta_{ik} \alpha \ $.

The dependence of the  density of dislocations flux tensor
on deviatoric stress is taken from \cite{Tsai}:
\begin{displaymath}
 j_{ik} = - \frac{S_{ik}}{\sqrt{J_2^S}} b N v \
\end{displaymath}
where $S_{ij}$ and $J_2^S$ are the elastic deviatoric stress
and it's second invariant, respectively:
\begin{displaymath}
 S_{ik} = \sigma^{el}_{ik} - \frac{1}{3} \delta_{ik} \sigma^{el}_{ll}, \quad
   J_2^S = \frac{1}{2} S_{ik} S_{ik}  \
\end{displaymath}
Using the generalization of the Orowan equation
\begin{displaymath}
\delta \varepsilon^{\mathrm{pl}}_{ik} =
         - \frac{1}{2} ( j_{ik} + j_{ki} ) \delta t
\end{displaymath}
one finally obtain
\begin{equation}
\label{or}
\frac{ d\varepsilon^{\mathrm{pl}}_{ik} }{ dt } =
          \frac{S_{ik}}{\sqrt{J_2^S}} b N v
\end{equation}
The equation for the total dislocation density and the dislocation velocity
are written as
\begin{displaymath}
\frac{dN}{dt} = K \sigma_{eff_N}^\lambda N v + \dot{N}_{bin}
\end{displaymath}
\begin{displaymath}
v = v_0 \sigma_{\mathrm{eff}}^m sign (\sigma_{\mathrm{eff}}) \exp {-\frac{Q_v}{kT}}
\end{displaymath}
where the term  $\dot{N}_{bin}$ accounts for the binary dislocation reactions and
$K, \lambda, v_0, m, Q_v $ are the material parameters.

Effective stress is defined as
\begin{displaymath}
\sigma_{\mathrm{eff}} =  |\sigma - \xi \mu b \sqrt{N}|
\end{displaymath}
$\sigma$ is applied elastic stress and $\xi$ is the strain hardening factor.
A hardening function (\ref{hard}) is considered as a probable
alternative.

The straightforward extension of the model is possible to account for
the dislocation evolution along each slip system. However more accurate
such model may appear, one should bear in mind that the number of the material
parameters will blow up. For example, in the Orowan equation (\ref{or})
$b N v$ is to be changed to the sum over all slip systems \cite{kalan}
\begin{displaymath}
\sum_i b_i N_i (v_d)_i
\end{displaymath}
while the effective stress for the $i^{th}$ slip system should be written as \cite{theod}
\begin{displaymath}
(\sigma_{\mathrm{eff}})_i = |\sigma_i - \xi_i\mu b_i \sqrt{N_i}|
-\mu b_i \sqrt{(\sum_j (\chi_{ij} N_j)}
\end{displaymath}
Thus it has been decided to start with the model for the evolution of the total
dislocation density.

As has been mention already, all the models of the strain relaxation
during the growth reviewed above (as well as most
examples of the AH model applications to the density evolution in the bulk crystals,
starting with the classical papers \cite{Brown,Muller}) are written as the equations
in the layer thickness instead of time as an evolutionary variable. In
other words, it is assumed that the dislocation motion is
instantly frozen and no relaxation occur in the part of the layer
that has been already grown. If it is probably acceptable for the growth
of III-V thin films, it is certainly not true for the SiGe epitaxial growth:
as experiments show, the growth interruption does not stop the relaxation process.
The model outlined briefly in this section is a true transient one similar to
the model used recently for the growth of bulk crystals \cite{zhu}.
Moreover, it allows the uniform treatment of the growth itself and annealing.

\newpage
\section*{}

\end{document}